\documentclass[12pt,preprint]{aastex}

\usepackage{graphicx}
\usepackage{epstopdf}

\shorttitle{Solar System objects as cosmic rays detectors}
\shortauthors{Privitera et al.}

\begin{document}


\title{Solar System objects as cosmic rays detectors}


\author{P. Privitera\altaffilmark{1,2,3}}
\affil{Department of Astronomy and Astrophysics, University of Chicago,
    Chicago, IL 60637}

\and 
\author{P. Motloch\altaffilmark{3}}
\affil{Department of Physics, University of Chicago, Chicago, IL 60637}
\email{priviter@kicp.uchicago.edu}

\altaffiltext{1}{Department of Physics, University of Chicago.}
\altaffiltext{2}{Enrico Fermi Institute, University of Chicago.}
\altaffiltext{3}{Kavli Institute for Cosmological Physics, University of Chicago.}


\begin{abstract}
In a recent Letter \citep{andrews}, Jupiter is presented as an efficient detector for Ultra-High Energy Cosmic Rays (UHECRs), through measurement by an Earth-orbiting satellite of gamma rays from UHECRs showers produced in Jupiter's atmosphere.  
We show that this result is incorrect, due to erroneous assumptions on the angular distribution of shower particles. We evaluated other Solar System objects as potential  targets for UHECRs detection, and found that the proposed technique is either not viable or not competitive with traditional ground-based UHECRs detectors.
\end{abstract}





\section{Jupiter as a cosmic ray detector}
\label{jupiter}
A recent Letter \citep{andrews} claims that a gamma-ray detector on an Earth-orbiting satellite, e.g. Fermi \citep{Fermi}, would be sensitive to Ultra-High  Energy Cosmic Rays (UHECRs) interacting in Jupiter's atmosphere. The satellite instrument would detect gamma rays from the shower of particles (electrons, positrons and photons) induced by UHECRs skimming Jupiter's atmosphere, which may emerge  from the atmosphere in a beam pointing to Earth. This result can be shown to be incorrect by a simple argument. 

For a UHECR shower to be detectable, the number of gamma rays, $n_\gamma$, reaching the satellite must be greater than one: 
\begin{equation}
n_\gamma = \frac{N_\gamma f_\gamma(\theta_b) A}{\pi (D\theta_b)^2} > 1 , 
\label{eq:ng1}
\end{equation}
 where $N_\gamma$ is the number of gamma rays in the shower at its exit point from the atmosphere, $f_\gamma(\theta_b)$ is the fraction of gamma rays emitted within a solid angle of half-angle $\theta_b$, $D$ is the distance between Earth and Jupiter ($\approx 5~\rm{A.U.}$) and $A$ is the area of the satellite detector ($\approx 1 ~\rm{m}^2$). 
For a $10^{21}$ eV proton shower exiting Jupiter's atmosphere at its maximum development, the number of electrons\footnote{For the sake of simplicity, in the following we will use the term electrons to indicate both electrons and positrons.} at the shower maximum is $N_e\approx 6\cdot10^{11}$ \citep{corsika,qgsjet}. The corresponding number of photons $N_\gamma$ was obtained by numerical integration of the solutions of the shower cascade equations \citep{rossi, lipari}. For a minimum gamma ray energy of 20~MeV (the lower detection limit in \citep{andrews}), we found  $N_\gamma = 1.3 \cdot N_e \approx 8\cdot10^{11}$. From Equation~\ref{eq:ng1}, we then derive: 
\begin{equation}
\frac{f_\gamma(\theta_b)}{\theta_b^2} > 2 \cdot 10^{12}  ~\rm{rad^{-2}}.  
\end{equation}
The inequality is fulfilled for $\theta_b<7\cdot10^{-7}$~rad and $f_\gamma(\theta_b)=1$ (or for an even smaller $\theta_b$ if $f_\gamma(\theta_b)<1$). Such an angular distribution would result in unphysical properties of the cosmic ray shower, for example a sub-cm radial extent incompatible by several order of magnitudes with the  characteristic size of a shower (e.g. Moliere radius in air $\approx80$~m). 
 Thus, we conclude that UHECR showers in Jupiter's atmosphere cannot be detected by an Earth-orbiting  satellite instrument. 
  
The results of \citep{andrews} are affected - in addition to other questionable approximations - by an erroneous assumption on the angular distribution of cosmic ray showers particles, namely that photons produced by an electron of energy $E_e$  have an average angle $\approx m_e / E_e$ with respect to the shower axis, where $m_e$ is the electron mass. In fact, $m_e / E_e$ is representative of the angular scales occurring in the elementary processes that drive the shower development (bremsstrahlung, Compton scattering and pair production). Instead, the angular distribution of both electrons and photons in a shower is dominated by multiple scattering of the electrons \citep{rossi} with a characteristic angle of $\approx E_s / E_e$, where $E_s=21$~MeV.  In the oversimplified treatment of \citep{andrews}, the flux of photons is thus overestimated by a factor $(E_s/m_e)^2$.   

\section{The Earth and other Solar System objects}
It may be interesting to evaluate other Solar System objects as potential cosmic ray detectors.  In the following, we analyze in more details UHECRs skimming the atmosphere of the closest object, Earth.    

\begin{figure}
\plotone{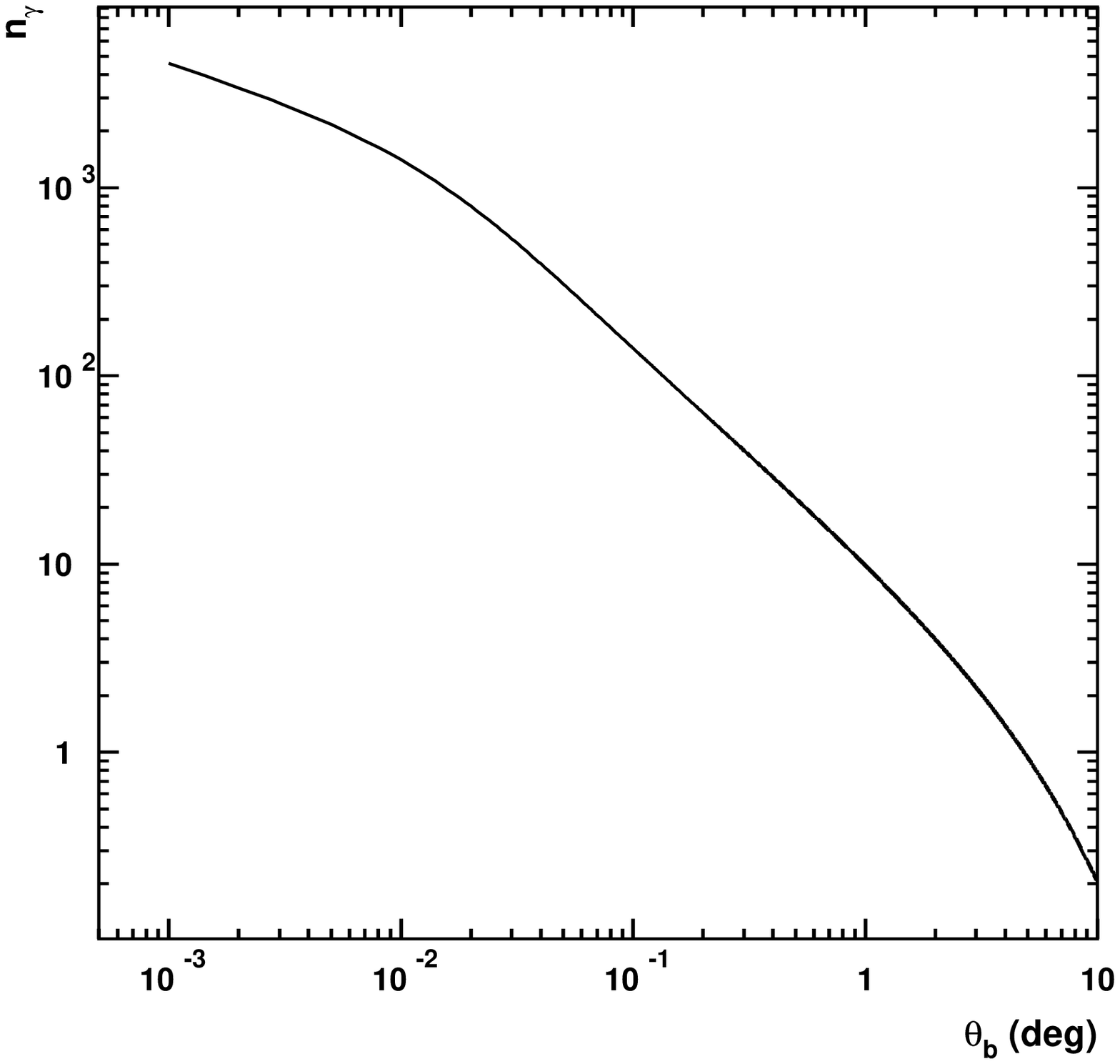}
\caption{Number of gamma rays detected by an Earth-orbiting satellite as a function of the angle of photon emission $\theta_b$.
 Gamma rays are produced by a $10^{21}$~eV UHECR shower skimming the Earth atmosphere. 
\label{fig:ngamma}}
\end{figure}

With a better approximation than Equation~\ref{eq:ng1}, the condition to detect a UHECR shower can be written as :
\begin{equation}
n_\gamma (\theta_b)= N_\gamma \frac{df_\gamma}{d\Omega}(\theta_b) \Delta\Omega > 1 , 
\label{eq:ng2}
\end{equation}
where $df_\gamma/d\Omega$ is the normalized photon angular distribution and $\Delta\Omega=A/D^2$ is the solid angle subtended by the satellite instrument. Detectable showers come from directions close to the satellite horizon, corresponding to a distance $D\approx2500$~km for an orbit of 500 km altitude. 
 The angular distribution of photons follows closely the angular distribution of electrons in the shower \citep{rossi}. 
To estimate $df_\gamma/d\Omega$, we convoluted the electron angular distribution, parametrized in  \citep{angular} as a function of the electron energy, with the electron energy spectrum at shower maximum as given in \citep{nerling}.  A minimum electron energy of 20~MeV was required for consistency with Section~\ref{jupiter}. Also, we took $N_\gamma \approx 8\cdot10^{11}$, corresponding to a $10^{21}$~eV UHECR (see Section~\ref{jupiter}).  
 In Figure~\ref{fig:ngamma}, $n_\gamma$ is shown as a function of the angle $\theta_b$. 
The detectability condition expressed by Equation~\ref{eq:ng2} is fulfilled up to a maximum angle $\theta_b^{\rm{max}} \approx 4^\circ$.

\begin{figure}
\plotone{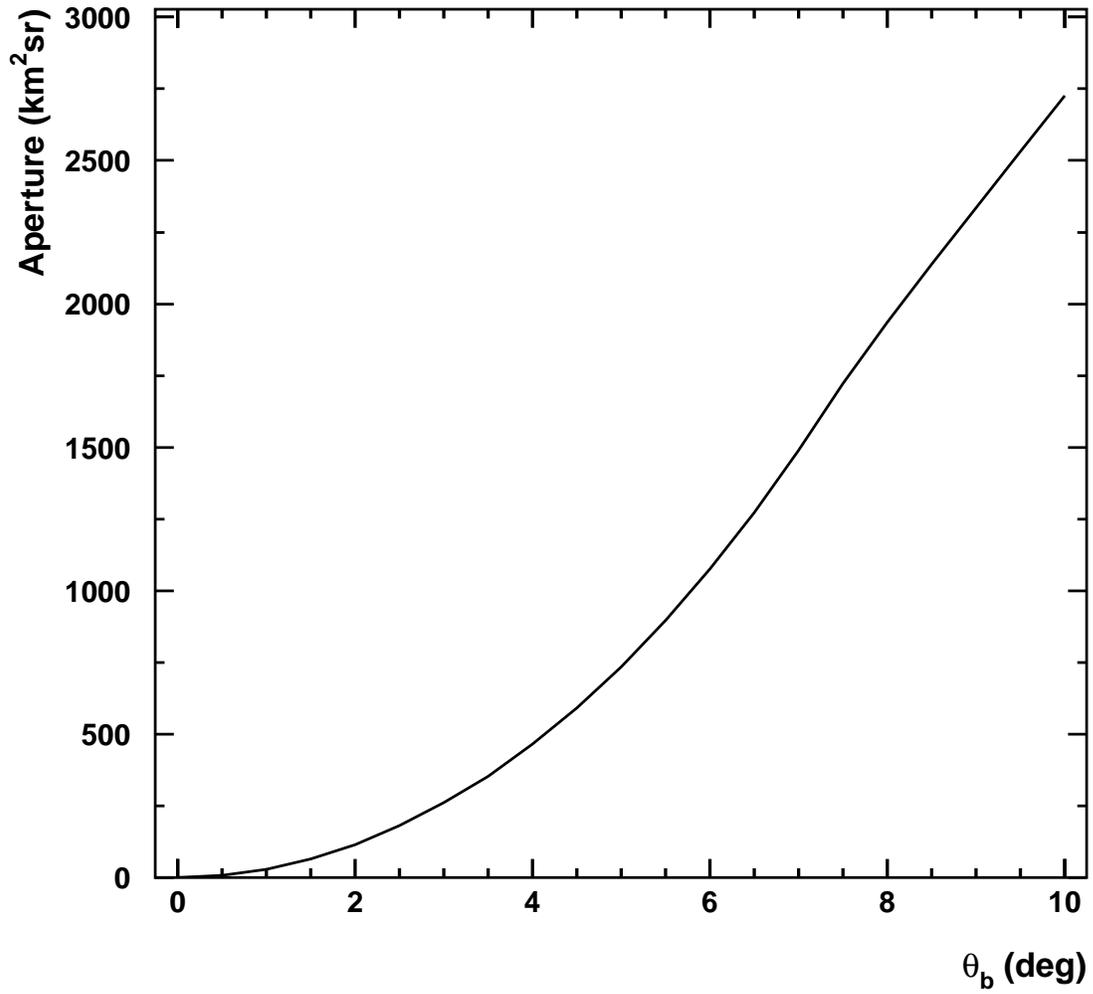}
\caption{Aperture of a satellite instrument detecting Earth-skimming UHECR showers, as a function of $\theta_b$.
\label{fig:aperture}}
\end{figure}

Thus, a UHECR shower skimming the Earth atmosphere is in principle detectable by a satellite instrument. However, it remains to be proven that a significant statistics of UHECRs may be collected with this technique. To estimate the aperture of an orbiting satellite, we performed a Monte Carlo simulation. A uniform distribution of the shower's impact point was generated over the Earth atmosphere taken as a spherical surface. The shower direction was then generated according to an isotropic distribution. For each shower, the column density along its path in the atmosphere was calculated using the US Standard Atmosphere model \citep{usatmo}. To ensure enough particles at the exit point, only showers with a column density between 600~$\rm{g/cm}^2$ and 1200~$\rm{g/cm}^2$ were further considered (within this range, $4 \cdot 10^{11} \le N_\gamma \le 8\cdot10^{11}$ for a $10^{21}$~eV proton shower). A shower was considered to be detected when the satellite was within an angle $\theta_b$ from the shower direction, with the vertex of the detection angle located at the point where the shower exits the atmosphere. In Figure~\ref{fig:aperture}, the geometrical aperture estimated from the simulation is shown as a function of $\theta_b$. The aperture for $\theta_b=\theta_b^{\rm{max}} \approx 4^\circ$ is found to be $\approx 500$ km$^2$sr. 
Lower UHECR energies will result in smaller apertures: for example, a $10^{19}$~eV shower would have $\theta_b^{\rm{max}} \approx 0.1^\circ$ (cf. Figure~\ref{fig:ngamma}) and a corresponding  aperture of only $\approx 0.2$ km$^2$sr (cf. Figure~\ref{fig:aperture}). 
For comparison, the geometrical aperture of the Pierre Auger Observatory \citep{auger}, the largest ground based UHECR facility, amounts to  $\approx7000$ km$^2$sr for all UHECR energies above $10^{18.5}$~eV. 

We also evaluated the case of UHECRs skimming the Moon surface, where the shower develops in the lunar regolith before exiting the surface. The corresponding aperture was found to be even smaller than that obtained for Earth.  Other Solar System objects were also excluded as effective cosmic ray detectors by similar calculations. 

Notice that the aperture in Figure~\ref{fig:aperture} is valid for any beamed emission from UHECR air showers. In particular, our findings also apply to radio emission from UHECR showers, which has similar angular scales \citep{radio}. 

\section{Conclusions}
We have critically reviewed the claim \citep{andrews} that Jupiter can be an efficient detector for UHECRs. We found that the number of gamma rays from a UHECR shower skimming Jupiter's atmosphere is too low to be detectable by an Earth-orbiting satellite. We also investigated the potential of such a technique for other Solar System objects. In the best case of Earth, we found that UHECRs skimming the Earth atmosphere are in principle detectable, but the aperture of an orbiting satellite is much smaller than traditional ground arrays. This result is also valid for beamed radio emission from UHECR showers. We conclude that this technique is not worthwhile further attention. 
 
\acknowledgments

This work was supported by the NSF grant PHY-1068696 at the University of Chicago, and the Kavli Institute for Cosmological Physics through grant NSF PHY-1125897 and an endowment from the Kavli Foundation.

\clearpage



\clearpage


\begin{thebibliography}{}

\bibitem[Abraham et al. 2004]{auger} Abraham, J., et al. [Pierre Auger Collaboration] 2004, Nucl. Instr. and Meth., A523, 50
\bibitem[Atwood et al. 2009]{Fermi} Atwood, W. B., et al. 2009, ApJ, 697, 1071
\bibitem[Heck et al. 1998]{corsika}Heck, D., Knapp, J., Capdevielle, J. N., Schatz, G. \& Thouw T. 1998, Forschungszentrum Karlsruhe Report FZK 6019
\bibitem[Kalmykov et al. 1997]{qgsjet}Kalmykov, N. N., S.S. Ostapchenko, S. S., \& Pavlov A. I. 1997, Nucl. Phys. B (Proc. Suppl.) 52B, 17 
\bibitem[Lafebre et al. 2009]{angular} Lafebre, S., Engel, R., Falcke, H., et al. 2009, Astropart. Phys., 31, 243 
\bibitem[Lipari 2009]{lipari} Lipari, P. 2009, Phys. Rev. D, 79, 063001
\bibitem[Motloch et al. 2014]{radio} Motloch, P., Hollon, N., \& Privitera, P. 2014, Astropart. Phys., 54, 40 
\bibitem[Nerling et al. 2006]{nerling} Nerling, F., Bl\"umer, J., Engel, \& R., Risse, M. 2006, Astropart. Phys., 24, 421 
\bibitem[Rimmer et al. 2014]{andrews} Rimmer, P. B., Stark, C. R., \& Helling, Ch., 2014,   ApJ, 787, L25
\bibitem[Rossi \& Greisen 1941]{rossi} Rossi, B., \& Greisen, K. 1941, Rev. Mod. Phys., 13, 240
\bibitem[U.S. Stand. Atm. 1976]{usatmo} U.S. Standard Atmosphere, 1976, U.S. Government Printing Office, Washington, D.C



\end{thebibliography}
\end{document}